\documentclass[12pt]{amsart}

\usepackage{amsaddr}
\usepackage[authoryear,round]{natbib}
\usepackage{float}
\usepackage{hyperref}
\usepackage[pdftex]{graphicx}
\usepackage{caption}
\usepackage[a4paper, margin=2.3cm]{geometry}
\usepackage{bm}

\theoremstyle{remark}

\makeatletter
\def\ex{\mathalpha{\operator@font E}}
\def\cov{\mathop{\operator@font Cov}\nolimits}
\def\var{\mathalpha{\operator@font Var}}
\makeatother

\begin{document}

\title{Effect Size Estimation in Linear Mixed Models}

\author{J\"{u}rgen Gro{\ss}}
\address{Institute for Mathematics and Applied Informatics, University of Hildesheim, Germany}
\email{juergen.gross@uni-hildesheim.de}
\author{Annette M\"{o}ller}
\address{Faculty of Business Administration and Economics,  Bielefeld University, Germany}
\email{annette.moeller@uni-bielefeld.de}
\thanks{Support of the second author by the Helmholtz Association’s pilot project ”Uncertainty Quantification” is gratefully acknowledged.}



\date{}

\begin{abstract} In this note, we reconsider Cohen's effect size measure $f^2$ under linear mixed models and demonstrate its application by employing an artificially generated data set. It is shown how $f^2$ can be computed with the statistical software environment {\sf R} using {\tt lme4} without the need for specification and computation of a coefficient of determination.
\end{abstract}

\maketitle

\markboth{J.~Gro{\ss} \& A. M\"{o}ller}{Effect Size Estimation in Linear Mixed Models}

\section{Introduction}\label{sec:intro}

In studies with a large number of observations, statistical testing procedures are prone to detect even minor departures from a null hypothesis yielding very small p-values. However, since significance does not automatically imply relevance, measures for the size of the effect associated with a possible rejection of the null hypothesis are often recommended as a useful tool, see e.g. \citet{wilkinson1999statistical}.

A well known effect size measure in a regression context when a quantitative variable $Y$ depends on independent regressors (quantitative and/or qualitative) is Cohen's $f^2$, see \citet[Chapt. 9]{cohen1988statistical}. Under multivariate normality this measure is strongly related to an $F$ test of a linear hypothesis that a subset $B$ of independent variables does not substantially contribute to the explanation of $Y$, given a set $A$ of independent regressors already included in the model. One may argue that even if the regression coefficients associated with variables $B$ significantly differ from zero, their contribution may be assessed as relevant when some meaningful size of their effect is measured.

In this light studies may additionally want to report $f^2$ values for each variable in a regression model given the others, see for example Tables 2 to 5 in \citet{taylor2020reactions}. With regard to the procedure of calculating $f^{2}$, the authors refer to \citet{selya2012practical} who introduced a practical method in relation with SAS\textsuperscript{\textregistered} software. This approach even carries over to linear mixed models, i.e. the case that some independent variables are associated with random effects rather than with fixed effects. Hence, the original approach by \citeauthor{cohen1988statistical} is generalized to a certain extent, involving the additional estimation of a variance-covariance matrix.

In the following we reconsider and discuss the generalization of $f^2$ to linear mixed models. We also point to an alternative computational procedure which turns out to be especially useful in context with the statistical software environment {\sf R} \citep{Rsoftware} and well known package {\tt lme4}, see \citet{R:lme4}, for fitting linear mixed models. Our explanations are illustrated on the basis of an artificially generated data set.

\section{Linear Mixed Model}\label{sec:lmm}

Consider a linear mixed model (LMM) described by
\begin{equation}\label{E1}
\bm{y} = \bm{X} \bm{\beta} + \bm{Z} \bm{u} + \bm{e}\; ,
\end{equation}
where $\bm{y}$ is an $n\times 1$ observable random vector. It is assumed that the $n\times p$ model matrix $\bm{X}$
of full column rank $p$ can be partitioned as
\begin{equation}\label{E2}
\bm{X} = (\bm{1}_{n} : \bm X_{1}: \bm{X}_{2})\; ,
\end{equation}
where $\bm{1}_{n}$ denotes the $n\times 1$ vector of ones, while the $n\times p_{1}$ and $n\times p_{2}$ matrices $\bm{X}_{1}$ and $\bm{X}_{2}$ contain the values of $p_{1}+ p_{2} = p-1$ regressors. The $p\times 1$ vector $\bm{\beta}$ comprises $p$ unknown parameters $\beta_{0}, \beta_{1}, \ldots, \beta_{p-1}$ addressed as fixed effects. The $n\times q$ matrix $\bm{Z}$ contains the values of independent variables associated with an $q\times 1$ vector $\bm{u}$ of unobservable random effects. It is assumed that $\bm{u}$ has expectation $\bm{0}_{q}$ (the $q\times 1$ vector of zeroes) and variance covariance matrix $\cov(\bm{u}) = \sigma^2 \bm{D}$ with unknown parameter $\sigma^{2} >0$ and $q\times q$ matrix $\bm{D}$, which may depend on further unknown parameters.
For the $n\times 1$ vector $\bm{e}$ of unobservable random errors it is assumed that $\ex(\bm{e})=\bm{0}_n$ and
$\cov(\bm{e}) = \sigma^2 \bm{T}$, where the $n\times n$ positive definite matrix $\bm{T}$ may also depend on unknown parameters. Moreover, the assumption $\cov(\bm{u}, \bm{e}) = \bm{0}_{q,n}$ (the $q\times n$ matrix of zeroes) implies
\begin{equation}\label{E3}
\cov(\bm{y}) =  \sigma^2 \bm V, \quad \bm{V} = \bm{Z} \bm{D} \bm{Z}^{T} + \bm{T}\; .
\end{equation}
Then the above model may also be represented by the triplet $\{\bm{y}, \bm{X}\bm{\beta}, \sigma^2 \bm{V}\}$ and can be considered as a special case of the general Gauss-Markov model, see e.g. \citet{gross2004general}. As a matter of fact, formulas useful under model \ref{E1}) carry over from  a classical regression model $\bm{Q}^{-1} \bm{y} = \bm{Q}^{-1} \bm{X} \bm{\beta} + \bm{\varepsilon}$ with $\ex(\bm{\varepsilon}) = \bm{0}_{n}$ and $\cov(\bm{\varepsilon}) = \sigma^2 \bm{I}_{n}$, see \citet[Sect 2.7]{christensen2020plane}. Here $\bm{Q}$ denotes some nonsingular matrix satisfying $\bm{Q}\bm{Q}^{T} = \bm{V}$. Let us assume for the moment that  $\bm{V}$ is completely known. Then
\begin{equation}\label{E4}
\widehat{\bm{\beta}} = (\bm{X}^{T} \bm{V}^{-1}\bm{X})^{-1} \bm{X}^{T} \bm{V}^{-1} \bm{y}
\end{equation}
is the best linear unbiased estimator for $\bm{\beta}$, and
\begin{equation}\label{E5}
\widehat{\sigma}^{2} =  \frac{1}{\nu} (\bm{y} - \bm{X}\widehat{\bm{\beta}})^{T} \bm{V}^{-1}(\bm{y} - \bm{X}\widehat{\bm{\beta}}),\quad \nu = n - p\; ,
\end{equation}
is the usual unbiased estimator for $\sigma^2$. Consider the linear hypothesis $H_{0}: \bm{R}\bm{\beta} =\bm{r}$ versus
$H_{1}: \bm{R}\bm{\beta} \not=\bm{r}$ for a given $r\times p$ matrix $\bm{R}$ of full row rank and a given $r\times 1$ vector $\bf{r}$. The corresponding $F$ statistic in model (\ref{E1}) is
\begin{equation}\label{E6}
F = \frac{(\bm{R}\widehat{\bm{\beta}} -\bm{r})^{T} (\bm{R} \bm{B} \bm{R}^{T})^{-1}(\bm{R}\widehat{\bm{\beta}} -\bm{r})}{r \widehat{\sigma}^2}
= \frac{(\bm{R}\widehat{\bm{\beta}} -\bm{r})^{T}(\bm{R} \bm{B} \bm{R}^{T})^{-1}(\bm{R}\widehat{\bm{\beta}} -\bm{r})}{(\bm{y} - \bm{X}\widehat{\bm{\beta}})^{T} \bm{V}^{-1}(\bm{y} - \bm{X}\widehat{\bm{\beta}})} \cdot \frac{\nu}{r}\; ,
\end{equation}
where
\begin{equation}\label{E7}
\cov(\widehat{\bm{\beta}}) = \sigma^2 \bm{B}, \quad \bm{B} = (\bm{X}^{T} \bm{V}^{-1} \bm{X})^{-1}\; .
\end{equation}
Under multivariate normality the statistic $F$ follows a
$F_{r,f}$ distribution provided $H_{0}$ holds true.

\subsection{Effect Size}

Suppose that we are interested in the effect of the independent variables represented by the model matrix $\bm{X}_{1}$, given the variables represented by $\bm{X}_{2}$. The corresponding linear hypothesis reads $\bm{R}_{1} \bm{\beta} = \bm{0}_{p_1}$ with $\bm{R}_{1} = (\bm{0}_{p_{1}}: \bm{I}_{p_{1}}:\bm{0}_{p_{1},p_{2}})$. Hence, by reasoning similar to Cohen (1988), an appropriate measure for the size of the effect based on the above $F$ statistic is provided by
\begin{equation}\label{E8}
f^2  = \frac{(\bm{R}_{1}\widehat{\bm{\beta}})^{T}(\bm{R}_{1} \bm{B}\bm{R}_{1}^{T})^{-1}(\bm{R}_{1}\widehat{\bm{\beta}})}{(\bm{y} - \bm{X}\widehat{\bm{\beta}})^{T} \bm{V}^{-1}(\bm{y} - \bm{X}\widehat{\bm{\beta}})}\;,
\end{equation}
thereby removing the factor $\nu/r$ from the $F$ statistic. This generalizes Cohen's $f^{2}$ in the sense that if $\bm{D} = \bm{0}_{q,q}$, then the unobservable random vector $\bm{u}$ equals $\bm{0}_{q}$ with probability one, the LMM reduces to the usual linear model of fixed effects only, and $f^2$ becomes identical to the measure provided by formula (9.2.1) in \citet{cohen1988statistical}.

We note that it is also possible to compute $f^2$ as
\begin{equation}\label{E9}
f^2 = \frac{R_{A,B}^2 - R_{A}^2}{1 - R_{A,B}^2}
\end{equation}
for appropriately defined coefficients of determination $R_{A,B}^{2}$ derived under the full model and $R_{A}^{2}$ derived under a reduced LMM assuming that variables represented by $\bm{X}_{1}$ are not present at all. Such a formula is the basis for the widely applied computational procedure suggested by \citet{selya2012practical}.

\subsection{Operational Effect Size}

Formula (\ref{E8}) for $f^2$ is only operational when $\bm{V}$ is completely known, a condition not met in practical applications. According to \citet{harville1977maximum} a simple LMM is the ordinary mixed and random effects ANOVA model, also referred to as traditional variance components model, see \citet[Chapt. 5]{christensen2019advanced}.
Under this model, the variance-covariance matrix of $\bm{y}$ depends on a total of $m$ variance components. The $n\times q$ matrix $\bm{Z}$ is partitioned as $\bm{Z} = (\bm{Z}_{1}: \cdots: \bm{Z}_{m-1})$, where each $n\times q_{i}$ matrix $\bm{Z}_{i}$ is the design matrix corresponding to a qualitative variable with a certain number of levels. In such a case one may assume
\begin{equation}
\bm{D} = \text{diag}\left[(\sigma_{1}^{2}/\sigma^2) \bm I_{q_{1}}, \ldots, (\sigma_{m-1}^{2}/\sigma^2) \bm I_{q_{m-1}}\right]
\end{equation}
where $\sigma^2 >0$ and $\sigma_{1}^{2}, \ldots \sigma_{m-1}^{2}\geq 0$ are $m$ unknown variance components. Then
\begin{equation}
\bm{V} = \bm{I}_{n} + \sum_{i=1}^{m-1} (\sigma_{i}^{2}/\sigma^2) \bm{Z}_{i} \bm{Z}_{i}^{T}\; .
\end{equation}
If $\sigma_{i}^2 = 0$ for some $i$, then the corresponding $q_{i}\times 1$ random effect vector $\bm{u}_{i}$ equals $\bm{0}_{q_{i}}$ with probability one. Such a model, and also more general ones, may be fitted in {\sf R} with the package
{\tt lme4}, see \cite{R:lme4}.  From the fitting procedure its is possible to obtain an estimate for the  variance-covariance matrix
\begin{equation}\label{E13}
\widehat{\cov}({\widehat{\bm{\beta}}}) = \widehat{\sigma}^2 \widehat{\bm{B}}
\end{equation}
of the estimated fixed effects parameter vector $\bm{\beta}$. Then a corresponding estimate for $f^2$ is given as
\begin{equation}\label{E10}
f^2  = \frac{1}{\nu}(\bm{R}_{1}\widehat{\bm{\beta}})^{T}(\bm{R}_{1}\widehat{\cov}({\widehat{\bm{\beta}}}) \bm{R}_{1}^{T})^{-1}(\bm{R}_{1}\widehat{\bm{\beta}})\; .
\end{equation}
where the factor $1/\nu$ is required to compensate for the usage of $\widehat{\sigma}^2$ in the estimated variance-covariance matrix (\ref{E13}).
The application of this formula is illustrated in the following section. The merit of formula (\ref{E10}) lies in the fact that it can be applied whenever an estimate (\ref{E13}) is available. But then, of course, the actual outcome also depends on the estimation procedure, implying that different estimation methods for (\ref{E13}) may also lead to (slightly) different actual values of (\ref{E10}). This, however, is not the topic of our discussion.  Also, when applying (\ref{E10}) there is no need for computing coefficients of determination. Nonetheless it is possible to define an operational version of (\ref{E9}) which corresponds to (\ref{E10}). This is demonstrated in Sect. \ref{sect:coeff}, where the $R^2$ measure discussed in \citet{edwards2008r2} is used with an operational version of the $F$ statistic obtained in the same light as (\ref{E10}) by employing the very same variance-covariance estimate (\ref{E13}). Again, employing alternative computational methods or alternative measures $R^2$ in linear mixed models, see also \citet{nakagawa2013general, nakagawa2017coefficient}, may result in different actual values of $f^2$.

\begin{figure}[hbt]
\centering
\includegraphics[width=25pc,angle=0]{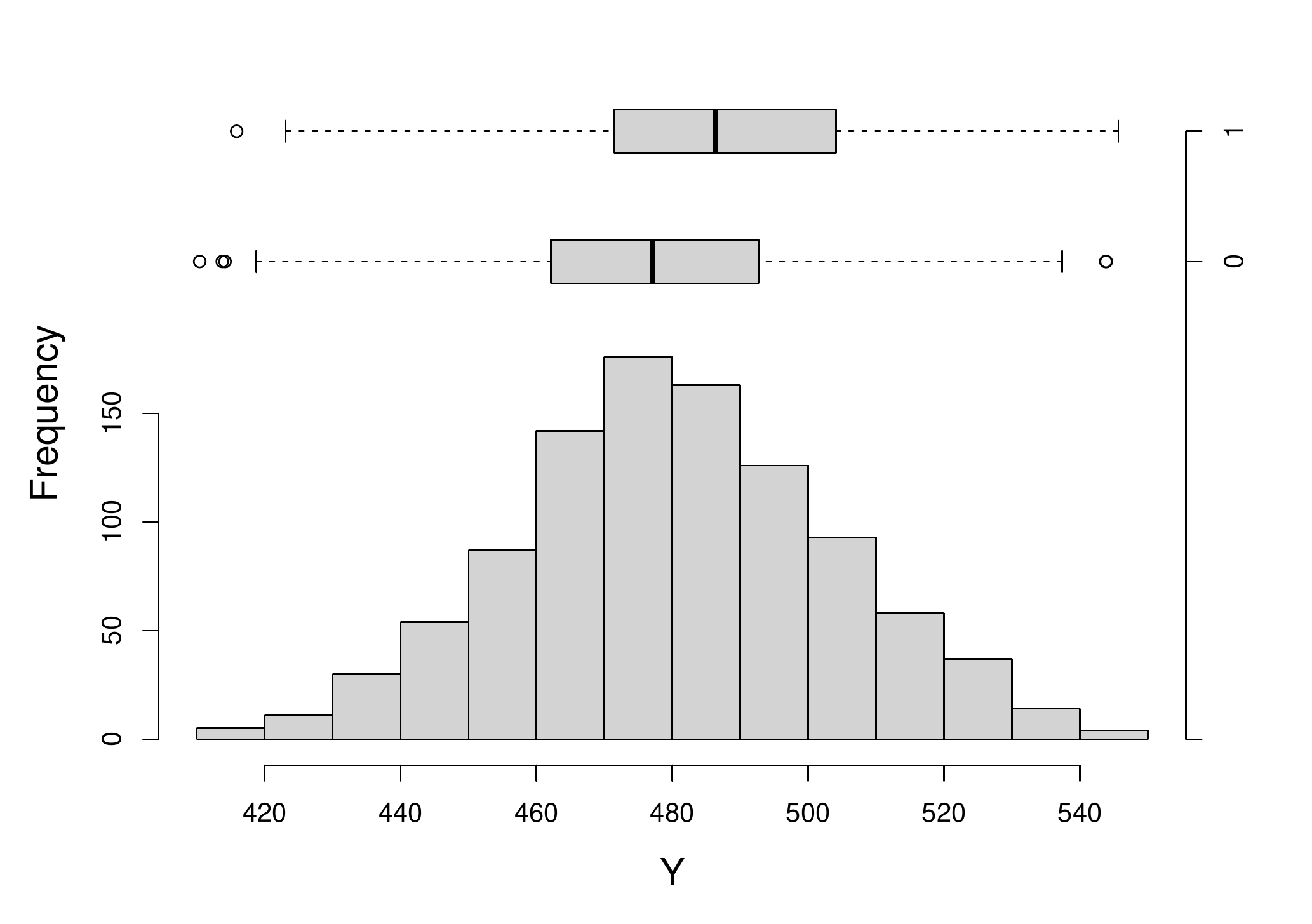}\\
  \caption{Distribution of response variable $Y$ over all $n=1000$ observations (histogram) and within two groups of sizes $n_{1} =687$ and $n_{2} = 313$ indicated by $0$ and $1$ (boxplots)}\label{F1}
\end{figure}

\section{Example}

In the following we discuss the performance of measure $f^2$ for an artificially generated data set of $n=1000$ observations intended to further illustrate some computational aspects. For the sake of simplicity our model consists of two independent variables $X_{1}$ (categorical/binary) and $X_{2}$ (quantitative) associated with fixed effects and one variable $Z$ (categorical) associated with random effects. However, the same principles apply when $X_{1}$, $X_{2}$, and $Z$ are extended to possible sets of variables containing more than one element. Our setting corresponds to the above mentioned variance components model with $p_{1}=p_{2}=1$ and $m=2$.

\subsection{Variable of Interest}

Figure \ref{F1} shows a discernible location difference with respect to the distribution of the response variable $Y$ in the two groups indicated by the binary variable $X_{1}$. The Welch two-sample $t$ statistic for the null hypothesis of no difference in group means reads  $|t|= 6.0751$ implying a highly significant result. A corresponding effect size measure is Cohen's $d$ which may be computed from {\sf R} package {\tt effectsize}, see \citet{R:effectsize}, as
$|d| = 0.4122$. From \citeauthor{cohen1988statistical}, values $|d|=0.2$, $|d|=0.5$ and $|d|=0.8$ indicate a small, medium and large effect, respectively.

\begin{figure}[hbt]
\centering
\includegraphics[width=25pc,angle=0]{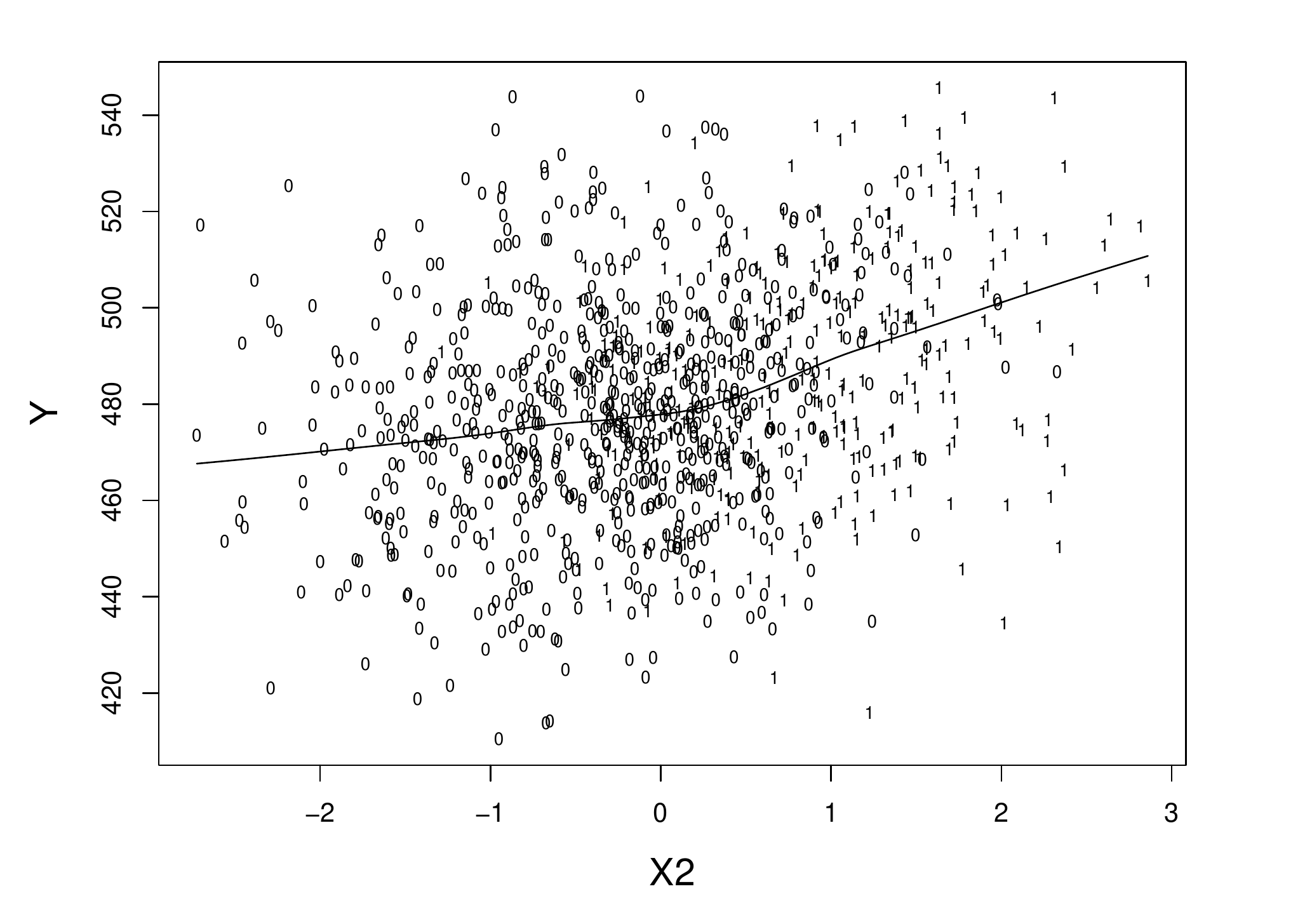}\\
  \caption{Scatter plot of $n=1000$ pairs $(X_{2}, Y)$ with group markings according to $X_{1}$}\label{F2}
\end{figure}

\subsection{Additional Fixed Effects}

From Figure \ref{F2} one may conclude, however, that the difference in the two groups may to some extent be explained by the variable $X_{2}$, since there is a tendency for larger values of $X_{2}$ to come along with larger values of $Y$ and observations from group 1 of variable $X_{1}$. Therefore one might be interested in the size of the effect of $X_{1}$ when $X_{2}$ is held constant. This can be achieved by considering a regression model with $X_{1}$ and $X_{2}$ as independent variables and deriving the measure $f^2$ as explained in \cite[Sect. 9]{cohen1988statistical}.  From package {\tt effectsize} one gets
$f^2 = 0.0017767$. Here, values $f^2=0.02$, $f^2=0.15$ and $f^2=0.35$ are supposed to indicate  a small, medium and large effect, respectively.

Recently, \citet{gross2023note} considered a generalized version $d_{\ast}$ of $d$ as an effect size measure for a binary variable $X_{1}$ given further variables. It may be computed from $f^2$ as
\begin{equation}
d_{\ast} = \sqrt{f^2 (n-2-w) \gamma}\; ,
\end{equation}
where in our analysis $w=1$ is the number of additional independent variables incorporated in the model  and $\sigma^2 \gamma$ is the variance of the regression coefficient for $X_{1}$. For our data $\gamma = 0.0065821$, yielding $d_{\ast} = 0.108$ and thus confirming a less than small effect.

\begin{figure}[hbt]
\centering
\includegraphics[width=25pc,angle=0]{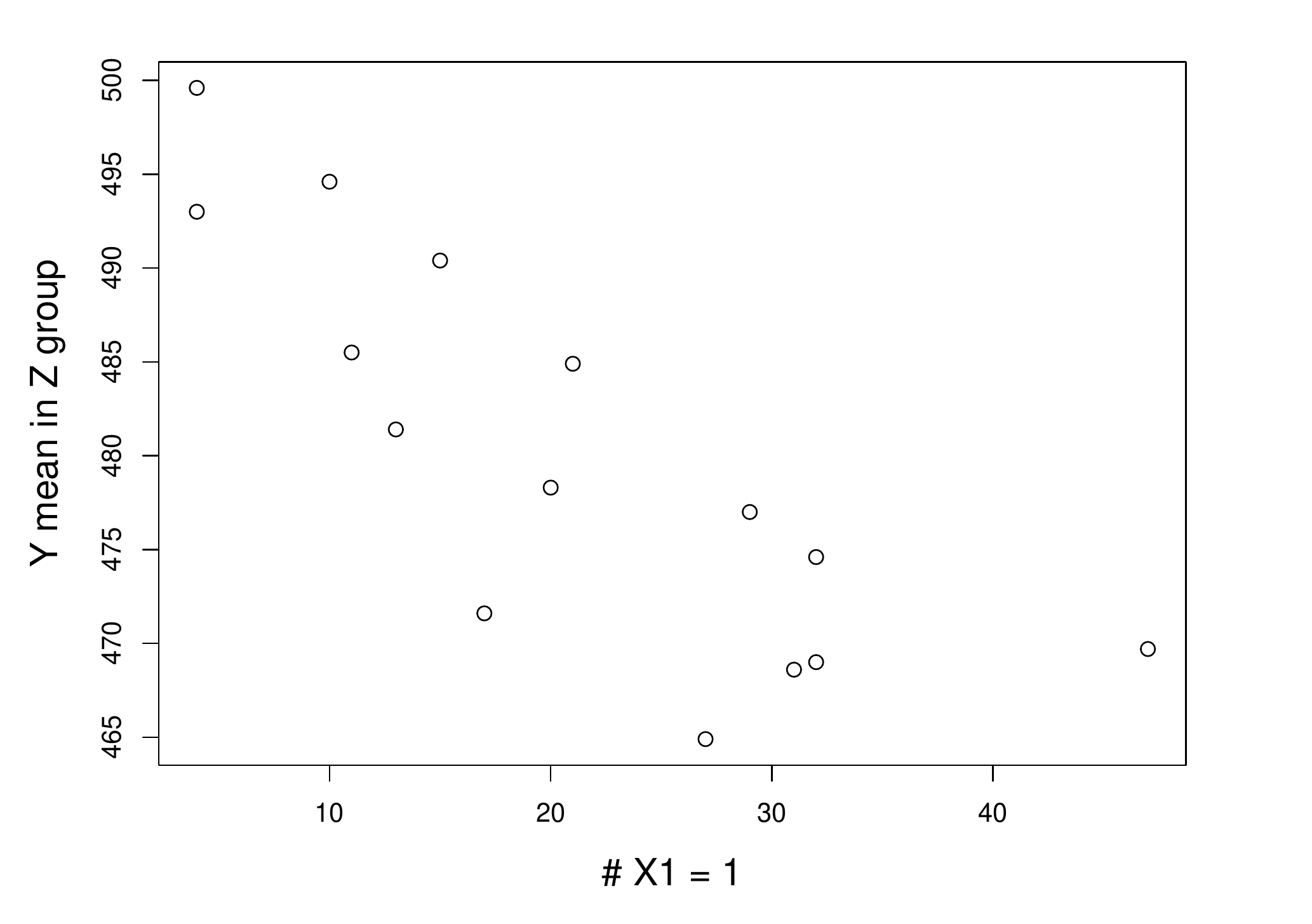}\\
  \caption{Scatterplot of $(\#\{X_{1} =1\}, \overline{Y})$ for the 15 groups of $Z$, where
   $\#\{X_{1} =1\}$ denotes the number of observations with $X_{1}=1$}\label{F3}
\end{figure}

\subsection{Additional Fixed and Random Effects}
As a next step one may take the categorical variable $Z$, admitting 15 groups in our data set, into account. From Figure \ref{F3} it is seen that there is a tendency for $Z$ groups with larger means of the response variable $Y$ to contain less observations marked as 1 (referring to variable $X_{1}$) than $Z$ groups with smaller means of $Y$. However, since observations from group 1 were earlier seen to reveal on average larger $Y$ values than observations from group 0, holding $Z$ constant is expected to contribute in favor of an effect of $X_{1}$ again.

\begin{figure}[hbt]
\centering
\includegraphics[width=25pc,angle=0]{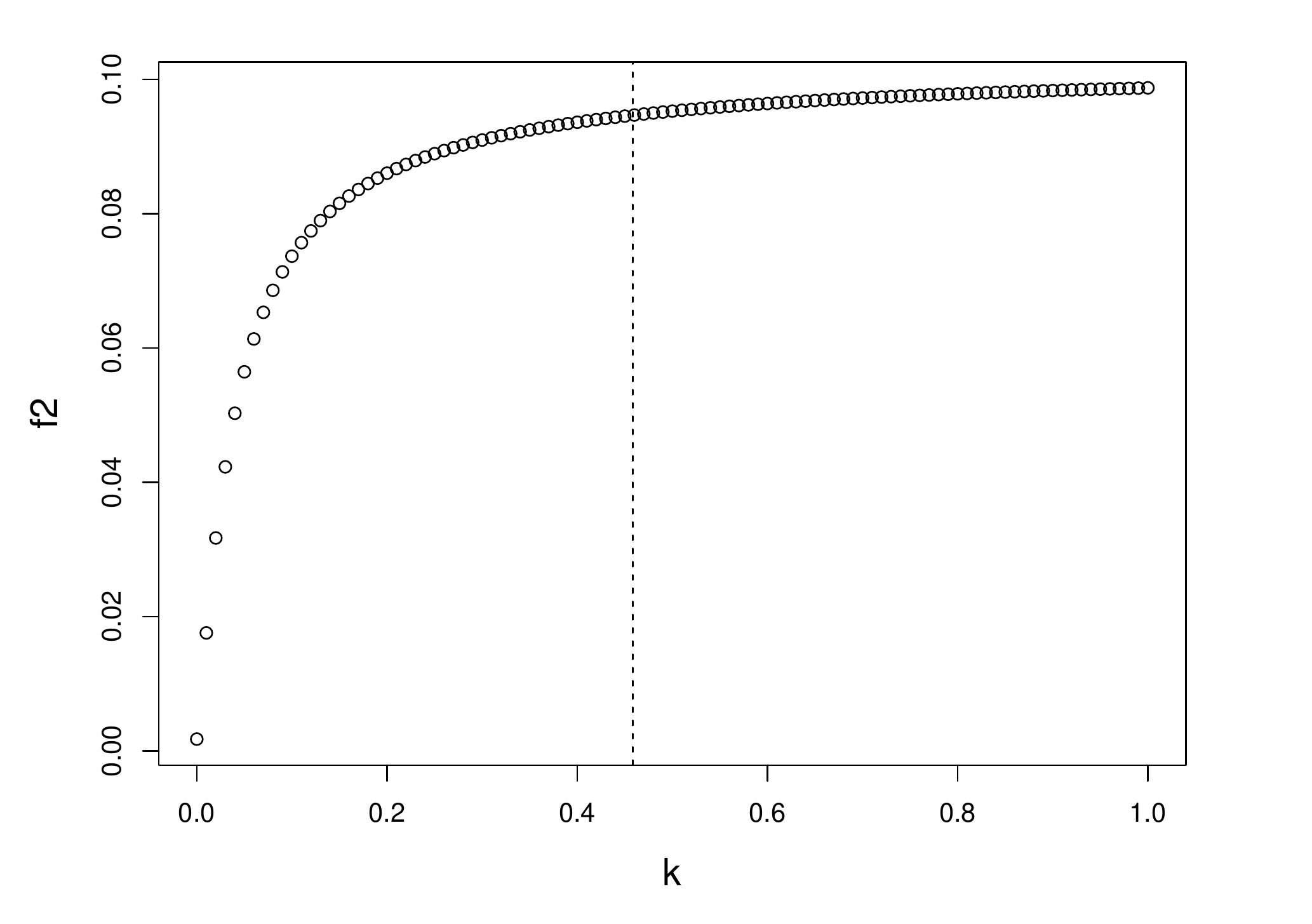}\\
  \caption{Values of $f^2$ from a LMM fit depending on $k$}\label{F4}
\end{figure}

As noted before, the variable $Z$ is meant to be associated with a random effects vector $\boldsymbol{u}$. The corresponding
design matrix $\bm{Z}$ has 15 columns where each row contains 0's and a single 1, indicating the membership of the observations to the respective $Z$ variable group. There are two variance components, the overall $\sigma^2 > 0$ and the random effects variance denoted by $\sigma_{u}^2\geq 0$ here. The model may also be reparameterized via an unknown $k\geq 0$ by setting $\sigma_{u}^2 = k \sigma^2$. This gives variance-covariance matrix
\begin{equation}\label{E11}
\cov(y) = \sigma^2 \bm{V}, \quad \bm{V} = k \bm{Z}\bm{Z}^{T} + \bm{I}_{m}\; ,
\end{equation}
and the LMM reduces to the usual fixed effects regression model in case $k=0$. For $\bm{V}$ from (\ref{E11}) one may
compute the effect size $f^2$  for $X_{1}$ as a function of $k$ from formula (\ref{E8}) when all variables $X_{1}$, $X_{2}$ and $Z$ are included in the LMM formulation. Figure \ref{F4} shows $f^2$ for different choices of $k$. As expected, the effect size is larger when both,  $X_{2}$ and $Z$ are incorporated into the model compared to the case when only $X_{2}$ is employed, corresponding to the choice $k=0$.

The operational version of $f^2$ from (\ref{E10}) can easily be computed from fitting the model by the function {\tt lmer} from package {\tt lme4} as
{\tt fit <- lmer(Y \~{} 1 + X1 + X2 + (1|Z))}. The estimated variance components are $\widehat{\sigma}^2 =  393.4455$ and $\sigma_{u}^2 = 180.4234$ giving
$\widehat{k} = \sigma_{u}^2/\widehat{\sigma}^2 = 0.4586$ as an estimation for $k$, see the dashed vertical line in Figure \ref{F4}. Then the vector $\widehat{\beta}$ is obtained from
{\tt fixef(fit)} and $\widehat{\cov}({\widehat{\bm{\beta}}})$ is obtained from {\tt vcov(fit)}. By using
$\bm{R}_{1} = (0,\, 1,\, 0)$ and $\nu= n-p = 997$, formula \ref{E10} yields
$f^2 = 0.0946626$ indicating a small but not medium effect size of $X_{1}$ when $X_{2}$ and $Z$ are held constant.

\subsection{Coefficient of Determination}\label{sect:coeff} Finally, we confirm that $f^2$ may also be obtained from formula (\ref{E9}), although there is no actual need for this when formula (\ref{E10}) can be used instead. For this, we define
\begin{equation}\label{E12}
R_{AB}^2 = \frac{(r/\nu)  F}{1 + (r/\nu)  F}, \quad r=p-1,\, \nu = n-p\; ,
\end{equation}
which is the proposed $R^2$ for linear mixed  models by \citet[Eq. (19)]{edwards2008r2}. Here, $F$ is the $F$ statistic from (\ref{E6}) for testing the hypothesis $H_{0}:\bm{R}\bm\beta = \bm{0}_{p-1}$ with $\bm{R} = (\bm{0}_{p-1}: \bm{I}_{p-1})$. The relationship  between $F$ and $R_{AB}^2$ from (\ref{E12}) may be established similar to  Example 4.8 from \citet{seber2003linear}. For our data $\bm{R} = (\bm{0}_{2}: \bm{I}_{2})$, $r= 2$, and $\nu = n-p= 997$. The measure $R_{A}^{2}$ is defined in the same way, but for a reduced model with all variables present except for $X_{1}$. For this reduced model we have $\bm{R} =(0, \, 1)$, $\bm{r} =(0)$, $r=1$, $\nu = n-p_{2}-1=998$. This results in
\begin{equation}
f^2 = \frac{R_{A,B}^2 - R_{A}^2}{1 - R_{A,B}^2}
= \frac{0.1539263 - 0.07418754}{1 - 0.1539263} = 0.09424569\; ,
\end{equation}
which is nearly the same as the value computed above. The displayed values were obtained from operational versions of $R_{A,B}^2$ and $R_{A}^2$ computed in the same light as formula (\ref{E10}) by employing
the estimated variance covariance matrix of the fixed effects resulting from applying {\tt vcov()} to the two models in question.

\end{document}